\documentclass[draft,11pt,showpacs,amsmath,amssymb]{revtex4}
\input{epsf}
\usepackage{graphicx} 
\usepackage{array}
\usepackage{dcolumn,longtable}

\usepackage{rotating,booktabs}
\usepackage{booktabs,threeparttable}

\def\rr{\mbox{\sf{}r}}

\begin{document}

\title{\bf
Calculations of polarizabilities and hyperpolarizabilities for the
Be$^{+}$ ion}
\author{Li-Yan Tang$^{1,2}$, Jun-Yi Zhang$^{3}$, Zong-Chao Yan$^{4,5}$, Ting-Yun Shi$^{1}$, James F. Babb$^{6}$
and J. Mitroy$^{3}$}

\affiliation {$^1$State Key Laboratory of Magnetic Resonance and
Atomic and Molecular Physics, Wuhan Institute of Physics and
Mathematics, Chinese Academy of Sciences, Wuhan 430071, P. R. China}

\affiliation {$^{2}$Graduate School of the Chinese Academy of
Sciences, Beijing 100049, P. R. China }

\affiliation {$^{3}$School of Engineering, Charles Darwin
University, Darwin NT 0909, Australia}

\affiliation {$^4$ Center for Cold Atom Physics, Chinese Academy of
Sciences, Wuhan 430071, P. R. China}

\affiliation{$^5$Department of Physics, Wuhan University, Wuhan
430072, China  and Department of Physics, University of New
Brunswick, Fredericton 4400, New Brunswick, Canada}

\affiliation{$^6$ITAMP, Harvard-Smithsonian Center for Astrophysics,
 60 Garden St., Cambridge, MA 02138, USA}
\date{\today}

\begin{abstract}
{The polarizabilities and hyperpolarizabilities of the Be$^+$ ion in
the $2\,^2\!S$ state and the $2\,^2\!P$ state are determined.
Calculations are performed using two
independent methods:  \textit{i)} variationally determined
wave functions using Hylleraas basis set expansions  and \textit{ii)}
single electron calculations utilizing a frozen-core
Hamiltonian. The first few parameters
in the long-range interaction potential between
a  Be$^+$ ion  and a
H, He, or Li atom, and the leading parameters of the effective potential for the
high-$L$ Rydberg states of beryllium were also computed. All the
values reported are the results of calculations close to convergence.
Comparisons are made with published results where available.}
\end{abstract}

\maketitle

\section{Introduction}

Studies of the Be$^+$ ion are of interest due to its importance in a
number of applications. First, the Be$^{+}$ ion is used as an
auxiliary ion to sympathetically cool other atomic or ionic
species~\cite{Rosenband,Rosenband2} that cannot be directly laser
cooled due to the lack of closed optical transitions.
Second, the Be$^+$ ion can combine with other atoms or neutral
molecules to form molecular ions, such as
BeH$^+$~\cite{Roth,Roth1,Roos,Errea}, and the study of the
long-range interaction between a Be$^+$ ion  and atoms or molecules
may open new routes for the study of state-elective chemical
reactions relevant to astrophysics~\cite{Roth}. Third,
investigations of Be$^+$ ion collisions with rare gases would be
useful in the study of ion-atom Feshbach
resonances~\cite{Idziaszek}, pressure broadening of
alkaline-earth-metal ions~\cite{Barklem}, and in studies of
excitation spectroscopy of the collision~\cite{Olsen,Andersen}.
Fourth, since beryllium has a number of isotopes, studies of the
Be$^+$ ion could potentially be used to determine the nuclear charge
radii of beryllium isotopes~\cite{Nortershauser}. Finally, there is
interest in studying the spectrum of the alkaline-earth-metal atom
in high angular momentum Rydberg states. Experimental investigations
have been made on a number of atoms~\cite{Snow1,Snow3,Snow5,Lyons}
with a view to determining the polarizabilities of the singly
ionized parent ion. These experiments measure the high $(n,L)$
energy splitting and then use a polarization model to extract the
polarizabilities. Recent calculations~\cite{Mitroy} have shown the
polarization model to be sensitive to nonadiabatic effects. The
Be$^+$ ion would be a useful candidate for a validation experiment
since it should be less sensitive to adiabatic effects and its
polarizabilities can be calculated to very high precision.

The above physical phenomena are influenced by the properties of the
Be$^{+}$ ion, and in particular the polarizabilities and
hyperpolarizabilities. There have been several calculations of the
Be$^+$ ion polarizabilities reported in the literature. These
include the work by Adelman and Szabo~\cite{Adelman} using the
Coulomb-like approximation, the calculation by Pipin and
Woznicki~\cite{Pipin} using the variation-perturbation approach with
a combined Hylleraas-configuration interaction (CI) basis set, the
calculation by Patil and Tang~\cite{Patil} using the valence
electron binding energy to construct wave functions constrained to
have the correct long-range asymptotic behavior, and finally the
large basis full core plus configuration interaction (FCCI)
calculations by Wang and
collaborators~\cite{Wang,Chen,Chung2,Qu,Qu1,Qu2}. However, there
have been no calculations reported on the polarizabilities and
hyperpolarizabilities for the Be$^{+}$ ion excited states.

In this paper, the polarizabilities, hyperpolarizabilities and some
long range ion-atom dispersion coefficients involving Be$^+$ ion are
computed with two independent methods. First, oscillator strengths
for many low-lying transitions are determined. Next, the
polarizabilities and hyperpolarizabilities for the $2\,^2\!S$ state
and $2\,^2\!P$ state of Be$^{+}$ ion are computed  variationally
using expansions of the wave functions in  Hylleraas bases. The same
set of long-range parameters are also computed using a fixed core
plus semi-empirical polarization potential to describe the valence
electron. The agreement between the two different calculations will
be seen to be excellent. The long-range dispersion interactions
between Be$^+$ ion and the H, He or  Li atoms are given and once
again the agreement between the two sets of calculations is
excellent. Furthermore, we compute all the parameters needed to
define a Be$^+$ polarization potential (including terms up to
$\rr^{-8}$) to describe the high-$L$ Rydberg states of beryllium,
where $\rr$ is the ion-electron distance. All results of this paper
are given in atomic units ($e=\hbar=m_e=1$).

\section{THEORY AND METHOD}\label {theory}

\subsection{Hylleraas variational method} \label{hylleraas}

The calculations for Be$^+$ ion are very similar in style to those
for Li~\cite{Yan,Tang}. In the center of mass frame, the nucleus is
taken as the reference particle $0$, with mass $m_0$ and charge
$q_0$, $r_i$ is  the electron-nucleus distance, and $i=1, 2, 3$. The
nonrelativistic Hamiltonian of this system can be written in the
form
\begin{eqnarray}
H_0 &=& -\sum_{i=1}^3 \frac{1}{2\mu_i}\nabla_i^2
-\frac{1}{m_0}\sum_{i> j\ge 1}^3\nabla_i\cdot\nabla_j
+q_0\sum_{i=1}^3\frac{q_i}{r_i} +\sum_{i> j\ge
1}^3\frac{q_iq_j}{r_{ij}}\,, \label {eq:t2}
\end{eqnarray}
where $r_{ij}=|\textbf{r}_i-\textbf{r}_j|$ is the distance between
electrons $i$ and $j$, $q_i$ are the charge of the three electrons,
and $\mu_i=m_im_0/(m_i+m_0)$ is the reduced mass between the $i$th
electron and the nucleus. In the present paper, all the calculations
are done in the infinite nuclear mass approximation.

Significant progress has  been made recently in variational
calculations for three electron systems by using multiple basis
sets in Hylleraas coordinates~\cite{yan1,Puchalski}. These have the
functional form,
\begin{eqnarray}
\phi = r_{1}^{j_{1}}r_{2}^{j_{2}}r_{3}^{j_{3}}r_{12}^{j_{12}}r_{23}^{j_{23}}r_{31}^{j_{31}}e^{-\alpha
r_{1}-\beta r_{2}-\gamma r_{3}}
\mathcal{Y}_{(\ell_{1}\ell_{2})\ell_{12},\ell_{3}}^{LM_{L}}(\hat{{\bf{r}}}_{1},\hat{{\bf{r}}}_{2},
\hat{{\bf{r}}}_{3})\chi(1,2,3)\,, \label {eq:t6}
\end{eqnarray}
where $\mathcal{Y}_{(\ell_{1}\ell_{2})\ell_{12},\ell_{3}}^{LM_{L}}$
is a vector-coupled product of spherical harmonics to form an
eigenstate of total angular momentum $L$ and projection $M_L$, and
$\chi(1,2,3)$ is the three-electron spin function. The variational
wave function is a linear combination of anti-symmetrized basis
functions $\phi$. With some truncations to avoid potential numerical
linear dependence, all terms in Eq.~(\ref{eq:t6}) are included such
that
\begin{equation}
j_1+j_2+j_3+j_{12}+j_{23}+j_{31}\leq\Omega\,,\label{eq:t7}
\end{equation}
where $\Omega$ is an integer, and the convergence for the energy
eigenvalue is studied by progressively increasing $\Omega$.

For the He atom, taking the nucleus as the  reference particle 0, the
electron is labeled as particle 1 and the other electron is labeled
as particle 2. The wave functions are expanded in terms of the
explicitly correlated basis set in Hylleraas coordinates:
\begin{equation}
\phi = r_1^ir_2^jr_{12}^ke^{-\alpha r_1-\beta
r_2}\mathcal{Y}_{\ell_1\ell_2}^{LM}(\hat{\textbf{r}}_1,\hat{\textbf{r}}_2)\,.\label{eq:t8}
\end{equation}

For the hydrogen atom, we use a basis set of form
\begin{equation}
\phi = r^\ell e^{-\beta r/2}L_n^{(2\ell+2)}(\beta r)\mathcal{Y}_{\ell
m}(\hat{\textbf{r}})\,,\label{eq:t10}
\end{equation}
where $L_n^{(2\ell+2)}(\beta r)$ is the generalized Laguerre
polynomial and the parameter $\beta$ is chosen to be
$\beta=2/(\ell+1)$. This basis set has been proven to be numerically
stable as the size of the basis set is increased.

\subsection{Single electron model} \label{HFCP}

The detailed description of the procedure used to construct the
frozen-core Hamiltonian and the semi-empirical polarization
potential can be found in previous works by Mitroy and
collaborators~\cite{Mitroy,MitroyA,Mitroy3}. Accordingly, only the
briefest description is given here.

Initially, a Hartree-Fock calculation of the Be$^+$ ion ground state
was performed. The core orbitals were then fixed, and a
semi-empirical core polarization potential was added to the
Hamiltonian. The core dipole polarizability was taken to be
0.0523~\cite{MitroyA}. The cutoff parameters  in the semi-empirical
core polarization potential were $\rho_0 = 0.941$, $\rho_1 = 0.895$,
$\rho_2 = 1.200$ with all other $\rho_L$ set to 1.00~\cite{MitroyA}.
These values were chosen to reproduce the binding energies of the
low-lying states.

The low-lying states and pseudo-continuum states were obtained by
diagonalizing the fixed core Hamiltonian in a large basis of Laguerre
type orbitals. Typically, there were about 50 orbitals for each value
of the valence angular momentum. This is large enough to eliminate
the basis as a significant source of error.

The multipole matrix elements and oscillator strengths were computed
with a modified transition operator~\cite{MitroyA}. The adjustable
parameter in the modified multiple operator was set to $\rho =
1.00$. Core excitations are included in the dispersion parameter
calculation. Oscillator strength distributions were constructed from
independent estimates of the core polarizabilities ($\alpha_2
= 0.01532$~\cite{MitroyA} and $\alpha_3 =
0.01125$~\cite{McEachran}). The calculations using this approach are
termed the Hartree-Fock plus core polarization (HFCP) model.

The dispersion coefficient calculations involving H and He used
matrix element lists that were generated using basis functions that
were similar (but not identical) in construction to
Eq.~(\ref{eq:t8}) and (\ref{eq:t10}). The matrix element lists
for the Li atom came from a one electron model as described in
Ref.~\cite{Zhang2}.

\subsection{Polarizabilities} \label{polarizabilities}

Using perturbation theory, the adiabatic long-range interaction
potential for an ion-atom system up to the $R^{-8}$ term can be
written~\cite{Zhang} as
\begin{eqnarray}
V_{ab}(R)=-\frac{1}{2}Q_a^2\sum_{\ell=1}^3\frac{\alpha_\ell^{(b)}}{R^{2\ell+2}}
-\sum_{n=3}^4\frac{C_{2n}}{R^{2n}}\,, \label {eq:t22}
\end{eqnarray}
where $Q_a=\sum_i Q_i$ is the total charge of the ion $a$,
$\alpha_\ell^{(b)}$ is the $2^\ell$-pole static polarizabilities for
the atom $b$, $R$ is the distance between the ion and the atom, and
the $C_{2n}$ parameters are the dispersion coefficients. The first
term in Eq.~(\ref{eq:t22}) is the polarization interaction, which
does not lead to a frequency shift between the different states of
the ion. The second term is the dispersion interaction, which can
lead to a frequency shift between two different ion states when the
ion is immersed in a buffer gas. From  Eq.~(\ref{eq:t22}), we  see
that the establishment of the interaction potential $V_{ab}(R)$
accurate to $R^{-8}$ requires the static polarizabilities
$\alpha_1$, $\alpha_2$, and $\alpha_3$ for the atom $b$, and the
dispersion coefficients $C_6$ and $C_8$ between ion $a$ and atom
$b$. Once we have obtained the oscillator strength spectra between
the ground states and the intermediate states for the H, He, Li, and
Be$^+$ ion systems, we can calculate the Be$^+$ ion
polarizabilities, and the dispersion coefficients for various
combinations of these particles. The detailed derivation of the
formulae for calculating the polarizabilities, hyperpolarizabilities
and dispersion coefficients in Hylleraas coordinates can be found in
the Ref.~\cite{Tang}.

For the high-$L$ Rydberg states of an atom or an ion, where a single
electron is in highly excited state but still moves in the field of
a spherically symmetric core, the polarization interaction between
the core and a single Rydberg electron leads to the effective
potential~\cite{Drachman1,Drachman2,Snow1,Snow3,Mitroy},
 \begin{eqnarray}
V_{\rm
eff}(\rr)=-\frac{A_4}{\rr^{4}}-\frac{A_6}{\rr^{6}}-\frac{A_7}{\rr^{7}}
-\frac{A_8}{\rr^8}-\frac{A_{8L}L(L+1)}{\rr^8}+\cdots\,,\label{eq:t23}
\end{eqnarray}
where $\rr$ is the radial coordinate of the Rydberg electron and the
coefficients $A_n$ are properties of the free ion core. Comparisons
with high precision variational calculations of the Rydberg states
for the few-body systems, He~\cite{Yan5}, Li~\cite{Chen}, and
H$_2$~\cite{Jacobson} have been made. This functional form has also
been used to analyze the fine structure spectrum of the Rydberg
states of neutral Mg and Ba, resulting in estimates of the dipole
polarizabilities of the alkali-earth-metal ions Mg$^+$ and Ba$^+$
ground states~\cite{Snow1,Snow5,Mitroy3}.

According to definitions given
previously~\cite{Mitroy3,Drachman1,Drake}, the leading coefficient
$A_4$ is half the size of the static dipole polarizability,
\begin{equation}
A_4=\frac{\alpha_1}{2}\,,\label{eq:t24}
\end{equation}
with
\begin{equation}
\alpha_{1}=\frac{8\pi}{9}\sum_{n}\frac{|\langle
n_{0}0\|T_1\|n1\rangle|^{2}}{E_{n}-E_{0}}\,.\label{eq:t25}
\end{equation}
The notation $|nL\rangle$ indicates the intermediate state with main
quantum number $n$ and angular momentum number $L$, and $T_1$ is the
dipole transition operator, which satisfies the general expression
for the $2^{\ell}$-pole transition operator in the center of mass
frame,
\begin{equation}
T_{\ell}=\sum_{i=1}^3 q_i r_i^{\ell} Y_{\ell 0} (\hat{\bf{r}}_i)\,.
\end{equation}

The next term $A_6$ is composed of two separate terms,
\begin{equation}
A_6=\frac{\alpha_2-6\beta_1}{2}\,,\label{eq:t26}
\end{equation}
where $\alpha_2$ is the quadrupole polarizability and $\beta_1$ is
the first order nonadiabatic correction to dipole polarizability.
They are defined as
\begin{equation}
\alpha_{2}=\frac{8\pi}{25}\sum_{n}\frac{|\langle
n_{0}0\|T_2\|n2\rangle|^{2}}{E_{n}-E_{0}}\,,\label{eq:t27}
\end{equation}
\begin{equation}
\beta_{1}=\frac{4\pi}{9}\sum_{n}\frac{|\langle
n_{0}0\|T_1\|n1\rangle|^{2}}{(E_{n}-E_{0})^{2}}\,.\label{eq:t28}
\end{equation}
The $\rr^{-7}$ term $A_7$ also comes from two parts, namely,
\begin{equation}
A_7=-\frac{5\eta+16q\delta_1}{10}\,,\label{eq:t29}
\end{equation}
where $q$ is the charge on the core. $\eta$ arises from third
order perturbation theory and it can be expressed as
\begin{eqnarray}
\eta &=& \frac{32\pi\sqrt{10\pi}}{225}\sum_{m,n}\frac{\langle
n_{0}0\|T_1\|m1\rangle \langle m1\|T_1\|n2\rangle \langle
n2\|T_2\|n_{0}0\rangle}{(E_{m}-E_{0})(E_{n}-E_{0})} \nonumber \\
&+&\frac{16\pi\sqrt{6\pi}}{135}\sum_{m,n}\frac{\langle
n_{0}0\|T_1\|m1\rangle \langle m1\|T_2\|n1\rangle \langle
n1\|T_1\|n_{0}0\rangle}{(E_{m}-E_{0})(E_{n}-E_{0})} \,,
\label{eq:t30}
\end{eqnarray}
and $\delta_{1}$ is the second order nonadiabatic correction to the
dipole polarization energy,
\begin{equation}
\delta_{1}=\frac{4\pi}{18}\sum_{n}\frac{|\langle
n_{0}0\|T_1\|n1\rangle|^{2}}{(E_{n}-E_{0})^{3}}\,.\label{eq:t31}
\end{equation}
Quite a few terms contribute to $A_8$,
\begin{equation}
A_8=\frac{\alpha_3-15\beta_2+\epsilon-\alpha_1\beta_1+72\delta_1}{2}\,.\label{eq:t32}
\end{equation}
The octupole polarizability $\alpha_3$ is computed by
\begin{equation}
\alpha_{3}=\frac{8\pi}{49}\sum_{n}\frac{|\langle
n_{0}0\|T_3\|n3\rangle|^{2}}{E_{n}-E_{0}}\,,\label{eq:t33}
\end{equation}
while $\beta_2$ comes from the first nonadiabatic correction part to
the quadrupole polarization energy, and is defined as
\begin{equation}
\beta_{2}=\frac{4\pi}{25}\sum_{n}\frac{|\langle
n_{0}0\|T_2\|n2\rangle|^{2}}{(E_{n}-E_{0})^{2}}\,.\label{eq:t34}
\end{equation}
The term $\epsilon$ is defined
\begin{eqnarray}
\epsilon&=&\frac{32\pi^{2}}{81}\sum_{m,n}\frac{|\langle
n_{0}0\|T_1\|m1\rangle \langle
m1\|T_1\|n0\rangle|^{2}}{(E_{m}-E_{0})^{2}(E_{n}-E_{0})} \nonumber \\
&+& \frac{64\pi^{2}}{405}\sum_{m,n}\frac{|\langle
n_{0}0\|T_1\|m1\rangle \langle m1\|T_1\|n2\rangle|^{2}
}{(E_{m}-E_{0})^{2}(E_{n}-E_{0})}\,.\label{eq:t35}
\end{eqnarray}
The hyperpolarizability $\gamma_0$ of the ground state, and
coefficients $\epsilon$, $\alpha_1$, $\beta_1$ are related by the
identity $\gamma_0 = 12(\epsilon - \alpha_1 \beta_1)$. The
last term $A_{8L}$ is nonadiabatic in origin and defined by
\begin{equation}
A_{8L}=\frac{18\delta_1}{5}\,.\label{eq:t36}
\end{equation}

For excited states, the working expressions for the evaluation of
the polarizabilities $\alpha_1$, $\alpha_1^T$, $\alpha_2$, and
$\alpha_3$ and hyperpolarizabilities $\gamma_0$ and $\gamma_2$ are
given in Tang {\em et al} \cite{Tang}. These expressions are quite
lengthy and not reproduced here.

\subsection{Dispersion interactions} \label{dispersion}

The dispersion interaction, between two atoms, $A$ and $B$, for the
Hylleraas wave functions was calculated from tabulated lists of
matrix elements using sum rules \cite{Tang}.   For the case of
$C_6$ involving two $S$-state atoms, one can write the sum
rule
\begin{equation}
C_6 = \frac{3}{2} \sum_{mn} \frac{f^A_{0m}f^B_{0n}}
{\Delta E^A_{m0} \Delta E^B_{n0}(\Delta E^A_{m0} + \Delta E^B_{n0})}
\label{C6def}
\end{equation}
The sum is over all states of $P^o$-symmetry. The absorption oscillator
strength, $f_{0m}$ ,for a dipole transition from $0 \to n$, with an
energy difference of $\Delta E_{m0} = E_{0} - E_{m}$, is most
conveniently defined in the present context \cite{Yan,MitroyA}
as
\begin{equation}
f_{0n} =  \frac {2 |\langle \psi_{0};L_{0} \parallel  r{\bf C}^{1}({\bf \hat{r}})
\parallel \psi_{n};L_{n} \rangle|^2 \Delta E_{n0}}
{3(2L_0+1)}  \ .
\label{fvaldef}
\end{equation}
In this expression for the HFCP method, $L_0$ is the ground state
orbital angular momentum, and $r{\bf C}^{1}({\bf
\hat{r}})=\sqrt{\frac{4\pi}{3}}r{\bf Y}^1(\bf{\hat{r}})$ is the
operator for a dipole transition just for one electron. Explicit
expressions for $C_8$, and $C_{10}$, and for other symmetries can be
found in \cite{Tang}.

The dispersion interaction calculations for the HFCP wave functions
utilized the completely general procedures outlined by Zhang and Mitroy
~\cite{Zhang3,Zhang4}.  These calculations utilize sum rules involving
lists of reduced matrix elements of the multipole operator $r^{k} {\bf C}^k$
multiplied by angular recoupling factors. They are effectively equivalent
to Eq.~(\ref{C6def}) and the expressions in \cite{Tang} despite the
differences how the calculations are actually carried out.

\section{ Results And Discussion}\label {results}

\subsection{The energies and oscillator strengths of Be$^+$ ion}\label {energies}
Table~\ref{tab:1} shows the convergence study for the
nonrelativistic energy of the Be$^{+}$ ion ground state as the size
of the Hylleraas basis set is enlarged. The ratio $R(\Omega)$ is
defined by
\begin{equation}
R(\Omega)=\frac{E(\Omega-1)-E(\Omega-2)}{E(\Omega)-E(\Omega-1)}\,.\label
{eq:t37}
\end{equation}
The extrapolation was done by assuming that the ratio would stay
constant around $R(\Omega)=4.839$. It is clear from
Table~\ref{tab:1} that the energy converges to high precision as the
number of terms is progressively increased. The final
nonrelativistic energy is accurate to about 11 significant figures
for the ground state. The uncertainty in the final energy is set to
be equal to the extrapolation correction. The nonrelativistic
theoretical energies for other low-lying states of Be$^+$ ion are
tabulated and compared with experimental binding energies in
Table~\ref{tab:2}. Binding energies from the HFCP calculations are
also listed. The Hylleraas binding energies are generally in good
agreement with the experimental binding energies. There are two
effects which change the binding energies, finite mass and
relativistic effect. The finite mass effect will be about 0.001\%
and will probably act to decrease the magnitude of the binding
energy. The largest discrepancy for the $2\,^{2}\!S$ state is due
entirely to relativistic effects (see Table III of ~\cite{yan1}),
which contributes about 0.007\% of the final value.

Table~\ref{tab:4} lists the oscillator strengths for a number of the
Be$^+$ ion dipole transitions involving low-lying states. The final
values for the Hylleraas calculations are obtained with an
extrapolation procedure similar to that for the energy. Once again
the uncertainty in the oscillator strength is assigned to be equal
to the magnitude of the extrapolation correction. It is evident from
Table~\ref{tab:4} that there is  excellent agreement between the
Hylleraas and HFCP calculations. The largest discrepancy between the
two calculations is only about 0.1$\%$ (for the $3\,^{2}\!P \to
3\,^{2}\!D$ transition). The oscillator strengths from the Hylleraas
calculation could be used to improve  the National Institute of
Standards and Technology (NIST) tabulations~\cite{wiese,martin}.

The Hylleraas oscillator strength for the $2\,^2\!S$ $\to$
$2\,^2\!P$ transition is accurate to about 7 significant figures and
is compatible with an earlier Hylleraas calculation by Yan~{\it et
al.}~\cite{Yan2}. Some earlier large scale ab-initio calculations
also gave oscillator strengths that are compatible with the present
Hylleraas calculation for this transition. These include the
multi-configuration Hartree-Fock (MCHF) calculation of
Godefroid~{\it et al.}~\cite{Godefroid} and the FCCI calculation of
Chung {\it et al}~\cite{Chung2}. While the present oscillator
strengths are reported with 7 significant digits, finite-mass and
relativistic effects that are not included in the calculation could
conceivably alter the oscillator strengths beginning at the fifth
digit.

There have also been some high precision oscillator strengths
reported for the $2\,^2\!P \to 3\,^2\!D$ transition. The MCHF value
of Godefroid~{\it et al.}~\cite{Godefroid} and the FCCI value of
Qu~{\it et al.}~\cite{Qu} agree with the Hylleraas calculation to
better than four digits. Qu~{\it et al.}~\cite{Qu1,Qu2} have also
reported CI calculations for the $2\,^2\!S \to 3\,^2\!P$ and
$3\,^2\!D \to 4\,^2\!F$ transition and again there is agreement to
better than four significant digits. The experimental oscillator
strengths listed in the Table were measured using beam-foil
spectroscopy \cite{andersen}. These oscillator strengths have only low
precision and cannot discriminate between the higher quality
theoretical estimates.

\subsection{Polarizabilities of the Be$^+$ ion}

The convergence properties of the static dipole polarizability
$\alpha_{1}$, and hyperpolarizability $\gamma_0$ for the Be$^{+}$
ion ground state are presented in Table~\ref{tab:5}. Both of them
have converged to five significant figures. The extrapolation was
done by assuming that the ratio between two successive differences
would stay constant as the basis size increased towards infinity.
The uncertainty in the final value is set equal to the magnitude of
the extrapolation correction from the explicitly calculated value
computed with the basis of largest dimension.

Table~\ref{tab:6} shows the convergence as a function of basis size
for the scalar and tensor dipole polarizabilities $\alpha_1$,
$\alpha_1^T$, and the hyperpolarizabilities $\gamma_{0}$,
$\gamma_{2}$ for the first excited state of Be$^{+}$ ion. The
intermediate sums in this case have contributions from doubly
excited unnatural parity states with $L^{\pi} = 1^e$ and $L^{\pi} =
2^o$ (e.g. the unnatural parity $1^e$ state has two $\ell = 1$
electrons coupled to a total angular momentum of $L = 1$). The
contribution from the unnatural parity states is usually small. For
example, the unnatural parity $P^{e}$ states contribution of
0.020616 to $\alpha_1$ is about 1\%. The scalar dipole
polarizability is converged to five significant digits, and the
tensor dipole polarizability $\alpha_1^T$ is converged to six
digits. The hyperpolarizabilities, $\gamma_0$ and $\gamma_2$ are
accurate to the six and five significant figures respectively. There
was no major numerical cancellation in the hyperpolarizabilities in
our calculation, in contrast to the situation that prevails
for the Li hyperpolarizability, which suffers severely from
cancellations in the different parts of the calculation~\cite{Tang}.

There have been a number of accurate calculations of the multipolar
polarizabilities for the Li atom in its ground or lowest excited
states~\cite{Zhang2,Safronova,Tang}. However, there have been fewer
polarizability calculations for the Be$^+$ ion in its ground and
lowest energy excited state. Table~\ref{tab:8} gives a comparison
between the present results and previous calculations for the static
polarizabilities of Be$^+$ ion in the $2\,^2\!S$ and $2\,^2\!P$
states. One of the most notable features of the Table is the very
good agreement between the Hylleraas and HFCP multipole
polarizabilities. The overall level of agreement is at the 0.1$\%$
level. The one exception was the static dipole polarizability of the
$2\,^2\!P$ state, but it should be noted that this polarizability is
small due to cancellation between different terms in the oscillator
strength sum. The polarizability of Be$^+$($2\,^2\!P$) state is
relatively small because the Be$^+$($2\,^2\!P \to 2\,^2\!S$)
oscillator strength is negative while all the other Be$^+$($2\,^2\!P
\to n\,^2\!L$) $f$-values are positive. The net effect of the
cancellations is a reduction in the overall size of the
polarizability by a factor of five.

The Coulomb approximation polarizability~\cite{Adelman} and the
asymptotically correct wave function polarizabilities~\cite{Patil}
achieve about 1-2$\%$ accuracy in $\alpha_1$ and $\alpha_2$. This is
noticeably worse than any of the other polarizabilities listed in
the Table. The older Hylleraas-type calculations by Pipin and
Woznicki~\cite{Pipin} used the variation-perturbation approach to
estimate the polarizabilities (as opposed to oscillator strength sum
rules). Their value of $\alpha_1 = 24.5$~\cite{Pipin} is compatible
with the present value but not nearly as precise as the Hylleraas
polarizability. The variation-perturbation calculations using the FCCI
wave function \cite{Wang,Chen} gave polarizabilities that agree with
the present Hylleraas polarizabilities to better than 0.01$\%$.

The present results for the Be$^{+}$($2\,^2\!P$) state are the only
results reported for the higher order polarizabilities and the
hyperpolarizabilities. Previously M\'{e}rawa and R\'{e}rat reported
the calculations for $\alpha_1$ and $\alpha_1^T$ for the
Be$^{+}$($2\,^2\!P$) state using the time-dependent gauge-invariant
method (TDGI)~\cite{merawa}, but the underlying structure model for
this approach is less accurate than the present calculations and we
do not include their numerical values in Table~\ref{tab:8}. The
overall level of agreement between the Hylleraas and HFCP
calculations is very impressive when it is considered that there are
significant numerical cancellations in the calculation of $\alpha_1$
that lead to a small value.

\subsection{Effective potential for beryllium Rydberg state}

Recently there have been a number of investigations of ion
polarizabilities based on the interpretation of resonant excitation
stark ionization spectroscopy
(RESIS)~\cite{Snow1,Mitroy,Lyons,Snow5}. The energy splitting of
adjacent Rydberg levels with $\Delta L = 1$ is used to determine the
parent ion polarizabilities. One recent finding has been an
increased appreciation of the importance of nonadiabatic and higher
order polarizability terms proportional to $\rr^{-7}$ and $\rr^{-8}$ in
the interpretation of the RESIS spectra.

Table~\ref{tab:11} summarizes all the parameters necessary to define
the polarization series given by Eq.~(\ref{eq:t23}) for the Be$^+$
ion ground state. The data are presented since the neutral beryllium
series represents an ideal system upon which to validate the
underlying assumptions used in the analysis of the RESIS experiment.
The nonadiabatic effects are strong [the nonadiabatic dipole
polarizability of $\beta_1 = -81.78175(1)$ dominates the quadrupole
polarizability of $\alpha_2 = 53.7659(2)$ in the evaluation of $A_6
= -218.4622(1)$] and all the ``Hylleraas'' polarizabilities listed
in Table~\ref{tab:11} would have an overall level of precision
better than 0.1$\%$. Although no experiment has been done, a RESIS
experiment upon neutral beryllium would provide a stringent test on
the ability of an analysis based on Eq.~(\ref{eq:t23}) to extract
polarizabilities from a typical RESIS spectrum.

\subsection{The long-range dispersion coefficients}\label {dispersion coefficients}

Table~\ref{tab:9} and Table~\ref{tab:10} list the long-range
dispersion coefficients for the Be$^+$ ion interacting with the H,
He, and Li atoms. Table~\ref{tab:9} lists dispersion coefficients
when both atoms or ions are in their ground states.
Table~\ref{tab:10} gives dispersion coefficients when one of the
systems in an excited state. All of the dispersion coefficients have
been calculated independently using the Hylleraas and HFCP wave
functions. Besides the wave functions, the procedures used to
combine the lists of matrix elements were completely independent. As
far as we know, the data listed in Tables~\ref{tab:9} and
Tables~\ref{tab:10} are the only dispersion coefficients published
for these systems.

The level of agreement between the two sets of $C_n$ values is
generally excellent. For example, the largest difference between any
of the dispersion constants listed in Table~\ref{tab:9} is only
0.06$\%$, occurring for the Be$^{+}$($2\,^2\!S$)--Li($2\,^2\!S$)
value of $C_8$.

The high level of agreement also occurs for the $C_n$ values listed
in Table~\ref{tab:10}, the only case of a greater than 1$\%$
difference occurring for the Be$^+$($2\,^2\!P$)--Li($2\,^2\!S$)
dimer. In this case, the roughly 3$\%$ disagreement occurs as a
result of the previously mentioned cancellations in the oscillator
strength sum for the Be$^+$($2\,^2\!P$) polarizability.  The net
effect of the cancellations is a reduction in the overall size of
the dispersion constants by a factor of about 100. For example, the
first term ($\lambda = 0$) of Eq.~(52) of Tang {\em et
al.}~\cite{Tang} was $-115.545$ while the second term( $\lambda=2$)
was 117.3969.

We do not list dispersion coefficients for the state combinations
that allow Penning or associative ionization (this occurs when the
excitation energy of one atom is sufficient to cause ionization in
the other atom). When this is possible, there is a singularity in
the energy denominator of the oscillator strength sum rules which
makes it problematic to achieve convergence.

All the values in Table~\ref{tab:9} and Table~\ref{tab:10} provide
an important benchmark for the accurate determination of the
interaction potentials between Be$^+$ ion and the H, He, or Li
atoms. The BeH$^+$ ion is one of the few molecular ions~\cite{Roth}
for which a potential curve could be computed with an explicitly
correlated wave function. Hence the present values for the Be$^+$--H
long-range interaction could be used to help construct a very
accurate global potential surface for this system.

\section{Conclusion}\label {conclusion}

Fully correlated Hylleraas variational wave functions have been used
to determine definitive values for the oscillator strengths,
polarizabilities and hyperpolarizabilities for the Be$^+$ ion
$2\,^2\!S$ ground state and the $2\,^2\!P$ excited state. The
Hylleraas results for the polarizabilities of the $2\,^2\!S$ state
improve the accuracy of previous values by more than one order of
magnitude. Complementary calculations using a semi-empirical method
have also been done. The high level of agreement between the two
calculations at the 0.1$\%$ level of precision attests to the
utility of carefully formulated effective potential approaches,
which can give good descriptions of atomic structure with low
computational expense.

The long-range dispersion coefficients for the Be$^+$ ion
interacting with a H, a He, or  a Li atom have been evaluated. The
polarizabilities and dispersion coefficients provide reliable
references for the description of ion-atom collisions involving
Be$^+$ ion and also for high precision calculations of the potential
curves between the Be$^+$ ion and atoms such as H, He or  Li.

In addition, all the parameters of the effective polarization
potential for the Be$^+$ ion up to the $\rr^{-8}$ term have been
obtained. These parameters are extremely useful in the description
of high-$L$ Rydberg states of beryllium, and could be used in
future experiments to determine the ionization potential of
beryllium and also to describe the fine structure of beryllium atom
Rydberg series.
In addition, the present calculations lay the foundation for the
further investigations of relativistic and QED effects on the
polarizabilities and other properties of the Be$^+$ ion.

\begin{acknowledgments}
This work was supported by NNSF of China under Grant No. 10674154
and by the National Basic Research Program of China under Grant No.
2005CB724508. Z.-C.Y. was supported by NSERC of Canada and by the
computing facilities of ACEnet, SHARCnet, WestGrid, and in Part by
the CAS/SAFEA International Partnership Program for Creative
Research Teams. J.F.B was supported in part by the U.S. NSF through
a grant for the Institute of Theoretical Atomic, Molecular and
Optical Physics at Harvard University and Smithsonian Astrophysical
Observatory. J.M. and J.Y.Z would like to thank the Wuhan Institute
of Physics and Mathematics for its hospitality during their visits.

\end{acknowledgments}


\begin{longtable}{llllllllllllllllll}
\caption{\label{tab:1}Convergence of the Hylleraas calculation of
the nonrelativistic energy (in atomic units)
for the $2\,^2\!S$ state of Be$^{+}$ ion.}\\
\hline\hline \multicolumn{1}{l}{$\Omega$} && \multicolumn{1}{l}{No.
of terms} && \multicolumn{1}{c} {$E$($\Omega$)} && \multicolumn{1}{l}
{$E$($\Omega$)$-$$E$($\Omega$-1)}&&
\multicolumn{1}{l} {R($\Omega$)}\\
\hline
8&&       1589&&    --14.324 763 166 358  &&--0.000 000 051 721  &&8.211\\
9&&       2625&&    --14.324 763 174 596  &&--0.000 000 008 238  &&6.278\\
10&&      4172&&    --14.324 763 176 309  &&--0.000 000 001 713  &&4.809\\
11&&      6412&&    --14.324 763 176 663  &&--0.000 000 000 354  &&4.839\\
Extrap.&&    &&     --14.324 763 176 736(73) &&                  &&\\
\hline \hline
\end{longtable}

\begin{longtable}{llllllllllllllllll}
\caption{\label{tab:2}Theoretical nonrelativistic energies and
experimental energies of the low-lying states for the Be$^{+}$ ion,
(in atomic units). The numerical uncertainty in the theoretical
energies are given in brackets. The experimental valence binding
energies are taken from the National Institute of Standards database
~\cite{ralchenko,stanley}. The ground-state energy for the Be$^{2+}$
ion was taken from Ref.~\cite{Drake1}.} \\
\hline \hline
\multicolumn{1}{c}{State}&&\multicolumn{1}{c}{Hylleraas
}&&\multicolumn{1}{c}{$E_{\textrm{Hylleraas}}$$-$$E$(Be$^{2+}$)
}&&\multicolumn{1}{c}{HFCP }&&\multicolumn{1}{c}{Experiment
}&&\multicolumn{1}{c}{Ref.~\cite{yan1}
}\\
\hline
$2\,^2\!S$        && --14.324763176736(73) &&--0.669196938312  &&--0.669250  &&--0.669247  &&--0.66924793(2)\\
$2\,^2\!P$        && --14.17933329329(24)  &&--0.52376705486   &&--0.523755  &&--0.523769  &&--0.52376988(2)\\
$3\,^2\!S$        && --13.9227892683(5)    &&--0.2672230298    &&--0.267189  &&--0.267233  &&--0.26723367(3)\\
$3\,^2\!P$        && --13.8851502898(5)    &&--0.2295840513    &&--0.229527  &&--0.229582 \\
$3\,^2\!D$        && --13.87805405934(36)  &&--0.22248782091   &&--0.222482  &&--0.222478 \\
$4\,^2\!S$        && --13.7987166133(8)    &&--0.1431503748    &&--0.143131  &&--0.143152 \\
$4\,^2\!F$        && --13.780581705614(80) &&--0.125015467190  &&--0.125015  &&--0.125008 \\
$5\,^2\!G$        && --13.735568352173(16) &&--0.080002113749  &&--0.080002  &&--0.079997 \\
\hline \hline
\end{longtable}

\begin{longtable}{llllllllllllllllll}
\caption{\label{tab:4}Dipole oscillator strengths for the
selected transitions of Be$^{+}$ ion.} \\
\hline \hline \multicolumn{1}{l}{Transition} &&\multicolumn{1}{l}
{Hylleraas} &&\multicolumn{1}{l}{HFCP} &&\multicolumn{1}{l}
{NIST~\cite{wiese,martin}}&&\multicolumn{1}{l}
{Exp.~\cite{andersen}}&&\multicolumn{1}{c} {Other theory}
\\ \hline
$2\,^2\!S \rightarrow 2\,^2\!P$ && 0.49806736(6)  && 0.4985  &&0.505  &&0.54(3)  &&0.498067381(25) Hylleraas~\cite{Yan2}  \\
                                &&                &&         &&       &&      && 0.49813 FCCI~\cite{Chung2}; 0.49807 MCHF~\cite{Godefroid}\\
$2\,^2\!S \rightarrow 3\,^2\!P$ && 0.08316525(18) && 0.0828  &&0.0804 &&      &&0.08136 FCCI~\cite{Qu2} \\
$2\,^2\!P \rightarrow 3\,^2\!S$ && 0.06434157(29) && 0.0643  &&0.0665 &&0.048(5)  &&            \\
$2\,^2\!P \rightarrow 4\,^2\!S$ && 0.01021583(30) && 0.0102  &&0.010  &&      && \\
$2\,^2\!P \rightarrow 3\,^2\!D$ && 0.6319828(11)  && 0.6321  &&0.652  &&      && 0.63199 MCHF~\cite{Godefroid}\\
                                &&                &&         &&       &&      &&0.63197 FCCI~\cite{Qu}\\
$3\,^2\!S \rightarrow 3\,^2\!P$ && 0.8297696(15)  && 0.8307  &&       &&      &&\\
$3\,^2\!P \rightarrow 3\,^2\!D$ && 0.08103350(17) && 0.0804  &&       &&      &&\\
$3\,^2\!P \rightarrow 4\,^2\!S$ && 0.1345245(13)  && 0.1346  &&       &&      &&\\
$3\,^2\!D \rightarrow 4\,^2\!F$ && 1.01460194(11) && 1.0146  && 1.01   &&0.66(3)  &&1.0146 FCCI~\cite{Qu1} \\
$4\,^2\!F \rightarrow 5\,^2\!G$ && 1.34537126(12) && 1.3453  &&   \\
\hline \hline
\end{longtable}

\begin{longtable}{llllllllllllllllll}
\caption{\label{tab:5}Convergence of the dipole polarizability
$\alpha_{1}$ and hyperpolarizability $\gamma_0$ for the Be$^{+}$ ion
ground state (in atomic units). The number of intermediate states of
a given angular momentum are denoted as $N_S$, $N_P$ and $N_D$.}
\\ \hline \hline
\multicolumn{1}{l}{($N_{S}$,$N_{P}$)}&&&
\multicolumn{1}{l}{$\alpha_{1}$} &&& \multicolumn{1}{l}{
($N_S,N_P,N_D$)} &&& \multicolumn{1}{l}{$\gamma_0$}
\\
\hline
(1589,1174)  &&&24.495 332  &&&(1589,1174,1174)  &&&--115 21.13 20 \\
(2625,2091)  &&&24.496 067  &&&(2625,2091,2091)  &&&--115 21.31 84  \\
(4172,3543)  &&&24.496 408  &&&(4172,3543,3543)  &&&--115 21.31 96  \\
(6412,5761)  &&&24.496 522  &&&(6412,5761,5761)  &&&--115 21.27 68 \\
Extrap.      &&&24.496 6(1) &&&Extrap.           &&&--115 21.30(3)\\
 \hline \hline
\end{longtable}

\begin{longtable}{llllllllllllllllll}
\caption{\label{tab:6}Convergence of the dipole polarizabilities
$\alpha_{1}$, $\alpha_{1}^{T}$ and the hyperpolarizabilities
$\gamma_0$, $\gamma_2$ for the Be$^+$ ion in the $2\,^2\!P$ state
(in atomic units). The number of natural parity intermediate states
of a given angular momentum are denoted as $N_S$, $N_P$, $N_D$ and
$N_F$. The number of unnatural parity intermediate states of a given
angular
momentum are denoted as $N_{P^{\prime}}$ and $N_{D^{\prime}}$.  } \\
\hline \hline
\multicolumn{1}{l}{($N_S$,$N_P$,$N_D$,$N_{P^{\prime}}$)}&&&
\multicolumn{1}{l}{$\alpha_{1}$}&&& \multicolumn{1}{l}{
$\alpha_{1}^{T}$} &&& \multicolumn{1}{l}{
($N_S$,$N_P$,$N_D$,$N_F$,$N_{P^{\prime}}$,$N_{D^{\prime}}$)}&&&
\multicolumn{1}{l}{$\gamma_{0}$} &&&
\multicolumn{1}{l}{$\gamma_{2}$}\\
\hline
(1589,1174,1174,1106)&&&2.024 6197  &&&5.856 054 26 &&&(1589,1174,1174,1248,1106,1428) &&&109 11.665 61  &&&--7372.0 881\\
(2625,2091,2091,2002)&&&2.024 7235  &&&5.856 019 68 &&&(2625,2091,2091,2307,2002,2640) &&&109 13.221 87  &&&--7373.4 188\\
(4172,3543,3543,3413)&&&2.024 7465  &&&5.856 014 25 &&&(4172,3543,3543,4051,3413,4587) &&&109 13.576 50  &&&--7373.6 683\\
(6412,5761,5761,3413)&&&2.024 7537  &&&5.856 013 46 &&&(6412,5761,5761,6806,3413,4587) &&&109 13.572 18  &&&--7373.5 698\\
Extrap.              &&&2.024 76(1) &&&5.856 012(1) &&&Extrap.                         &&&109 13.57(1)   &&&--7373.6 1(5)\\
 \hline \hline
\end{longtable}

\begin{longtable}{lllllllllllllllllllll}
\caption{\label{tab:8}Comparisons of the static polarizabilities and
hyperpolarizabilities (in atomic units) for the $2\,^2\!S$ and
$2\,^2\!P$ states of Be$^{+}$ ion.}
\\ \hline \hline
&&&$2\,^2\!S$ state&&\\
 \multicolumn{1}{l}{Method}&
\multicolumn{1}{l}{$\alpha_{1}$} & \multicolumn{1}{l}{$\alpha_{2}$}
&\multicolumn{1}{l}{$\alpha_{3}$}
&\multicolumn{1}{l}{$\gamma_{0}$}\\
\hline
Coulomb approximation~\cite{Adelman} &24.77&&&\\
Variation-perturbation Hylleraas CI~\cite{Pipin}&24.5\\
Asymptotic correct wave function~\cite{Patil} &24.91&53.01&465.7  \\
Variation-perturbation FCCI~\cite{Wang,Chen} &24.495 &53.774(24) &465.79(11)\\
HFCP &24.493  & 53.760   & 465.77  &--11511\\
Hylleraas & 24.4966(1) &  53.7659(2) &  465.7621(1)  & --11521.30(3)\\
\hline
&&&$2\,^2\!P$ state&&\\
\multicolumn{1}{l}{Method}& \multicolumn{1}{l}{$\alpha_{1}$} &
\multicolumn{1}{l}{$\alpha_{1}^{T}$} &
\multicolumn{1}{l}{$\alpha_{2}$}  &\multicolumn{1}{l}{$\alpha_{3}$}
 &\multicolumn{1}{l}{$\gamma_{0}$} &\multicolumn{1}{l}{$\gamma_{2}$}\\
\hline
HFCP            &2.028 &5.835 &62.313 &1208.8 &10996  &--7450.4 \\
Hylleraas       &2.02476(1) &5.856012(1) &62.2840(1)
&1207.812(2) &10913.57(1) &--7373.61(5)\\
 \hline \hline
\end{longtable}

\begin{longtable}{llllllllllllllllll}
\caption{\label{tab:11}The polarizability parameters and
coefficients $A_n$ of the polarization potential
Eq.~(\ref{eq:t23}).}
\\ \hline \hline \multicolumn{1}{l}{Method} &&
\multicolumn{1}{l}{$\alpha_{1}$} && \multicolumn{1}{l}{$\beta_{1}$}
&&\multicolumn{1}{l}{$\alpha_{2}$} &&
\multicolumn{1}{l}{$\beta_{2}$} && \multicolumn{1}{l}{$\alpha_{3}$}
&& \multicolumn{1}{l}{$\delta_{1}$} && \multicolumn{1}{l}{$\eta$} &&
\multicolumn{1}{l}{$\epsilon$} \\ \hline
Hylleraas     &&24.4966(1) &&81.78175(1) &&53.7659(2) &&58.1169(1) &&465.7621(1) &&279.16401(2) &&917.569(1) &&1043.27(1)\\
HFCP          &&24.494      &&81.751       &&53.760      &&58.106      &&465.77       &&278.94        &&917.37      &&1038.7 \\
\hline
                &&$A_4$      && $A_6$       && $A_7$      && $A_8$      && $A_{8L}$\\
Hylleraas       &&12.24830(5) && --218.4622(1) && --905.445(2) && 9366.857(1) && 1004.99051(8) \\
HFCP            &&12.247 && --218.37 && --904.99 && 9357.2 && 1004.2 \\
 \hline \hline
\end{longtable}

\begin{longtable}{lll}
\caption{\label{tab:9}The long-range dispersion coefficients $C_6$
and $C_8$ for the Be$^{+}$ ion ground state interacting with the H,
He, and Li atom. The first row for each system came from Hylleraas
wave functions while the second row were computed
with HFCP wave functions.}  \\
\hline\hline
\multicolumn{1}{l}{System} & \multicolumn{1}{l}{$C_6$}
& \multicolumn{1}{l}{$C_8$}  \\
\hline
Be$^{+}$($2\,^2\!S$)-H($1\,^2\!S$)            & 18.8314(1)      &371.675(5)    \\
                                              & 18.829          &371.62        \\
Be$^{+}$($2\,^2\!S$)-He($1\,^1\!S$)           & 6.9811(1)       &120.425(3)   \\
                                              & 6.979           &120.44       \\
Be$^{+}$($2\,^2\!S$)-He($2\,^1\!S$)           & 621.577(2)      &41371.9(1)  \\
                                              & 621.52          &41370    \\
Be$^{+}$($2\,^2\!S$)-He($2\,^3\!S$)           & 400.289(3)      &19753.9(2)   \\
                                              & 400.26          &19753    \\
Be$^{+}$($2\,^2\!S$)-Li($2\,^2\!S$)           & 286.75(1)       &11991.1(2)    \\
                                              & 286.82          &11998           \\
\hline\hline
\end{longtable}


\begin{longtable}{llll}
\caption{\label{tab:10}The long-range dispersion coefficients $C_6$,
and $C_8$ for the Be$^{+}$ ion interacting with H, He, and Li atom
for the atomic states with a combined angular momentum of $L = 1$.
The $M_b$ column denotes the total magnetic quantum number of the
system. The first row for each system came from Hylleraas wave
functions while the second row were computed
with HFCP wave functions.} \\
\hline \hline System & $M_b$ & $C_6$ & $C_8$  \\
\hline
Be$^{+}$($2\,^2\!P$)-H($1\,^2\!S$)   & 0           & 38.53656(1)    &1291.81(1) \\
                                     & 0           & 38.571         &1292.7      \\
Be$^{+}$($2\,^2\!P$)-H($1\,^2\!S$)   & $\pm 1$     & 17.12124(1)    &172.578(1) \\
                                     & $\pm 1$     & 17.128         &172.67     \\
Be$^{+}$($2\,^2\!P$)-He($1\,^1\!S$)  & 0           & 11.64784(1)    & 438.3026(6) \\
                                     & 0           & 11.660         & 438.63  \\
Be$^{+}$($2\,^2\!P$)-He($1\,^1\!S$)  & $\pm 1$     & 6.028208(7)    & 34.0606(3) \\
                                     & $\pm 1$     & 6.030          & 34.113      \\
Be$^{+}$($2\,^2\!S$)-He($2\,^3\!P$)  & 0           & 791.311(4)      & 93960.4(9)  \\
                                     & 0           & 791.52          & 93960  \\
Be$^{+}$($2\,^2\!S$)-He($2\,^3\!P$)  & $\pm 1$     & 373.997(3)      & 3547.53(8)  \\
                                     & $\pm 1$     & 374.25          & 3553.1   \\
Be$^{+}$($2\,^2\!P$)-Li($2\,^2\!S$)  & 0           & --326.06(8)    &--3.613643(6)$\times 10^5$ \\
                                     & 0           & --325.74       &--3.52050$\times 10^5$   \\
Be$^{+}$($2\,^2\!P$)-Li($2\,^2\!S$)  & $\pm 1$     & 2.32(4)        &--1.214626(3)$\times 10^5$ \\
                                     & $\pm 1$     & 2.3964         &--1.1836$\times 10^5$  \\
Be$^{+}$($2\,^2\!S$)-Li($2\,^2\!P$)  & 0           & 925.279(2)     &1.043315(3)$\times 10^5$ \\
                                     & 0           & 925.49         &1.0438$\times 10^5$     \\
Be$^{+}$($2\,^2\!S$)-Li($2\,^2\!P$)  & $\pm 1$     & 420.470(1)     &3804.79(4)\\
                                     & $\pm 1$     & 420.53         &3806.7    \\
\hline\hline
\end{longtable}

\end{document}